\documentclass[aps,prl,twocolumn,showpacs,groupaddress]{revtex4}
\usepackage{amssymb}
\usepackage{graphicx}
\usepackage{dcolumn}
\usepackage{bm}
\usepackage{amsmath}

\begin{document}

\title{Two photon absorption and coherent control with broadband down-converted light}
\author{Barak Dayan}
\email[]{Barak.Dayan@Weizmann.ac.il}
\author{Avi Pe'er}
\email[]{Avi.Peer@Weizmann.ac.il}
\author{Asher A. Friesem}
\author{Yaron Silberberg}
\homepage[]{www.weizmann.ac.il/home/feyaron/}
\affiliation{Department of Physics of Complex Systems, Weizmann Institute of Science,\\
Rehovot 76100, Israel}

\begin{abstract}

We experimentally demonstrate two-photon absorption (TPA) with
broadband down-converted light (squeezed vacuum). Although
incoherent and exhibiting the statistics of a thermal noise,
broadband down-converted light can induce TPA with the same sharp
temporal behavior as femtosecond pulses, while exhibiting the high
spectral resolution of the narrowband pump laser. Using
pulse-shaping methods, we coherently control TPA in Rubidium,
demonstrating spectral and temporal resolutions that are 3-5
orders of magnitude below the actual bandwidth and temporal
duration of the light itself. Such properties can be exploited in
various applications such as spread-spectrum optical
communications, tomography and nonlinear microscopy.
\end{abstract}


\maketitle

\textfloatsep0.4cm

In two-photon absorption (TPA), two photons whose sum energy
equals that of an atomic transition, must arrive at the atom
together. The two, seemingly contradicting, demands of a narrow
temporal and a narrow spectral behavior of the inducing light are
typically maximized by transform-limited pulses, which exhibit the
highest peak intensity possible for a given spectral bandwidth.
Nonetheless, it was shown
\cite{Nature396,PRA60,Weiner&Zheng2,PRL1} that pulses can be
shaped in a way that will stretch them temporally yet will not
affect the transition probability, and even increase it in certain
cases. Other experiments have exploited coherent control
\cite{Tannor&Rice,Shapiro&Brumer,Warren&Rabitz,Rice&Zhao} to
increase the spectral selectivity of nonlinear interactions
induced by ultrashort pulses
\cite{Lang&Mutzkus,Nirit&Dan,Gershgoren&Bartles}; however, the
spectral resolution demonstrated by these methods remains
considerably inferior to that obtained by narrowband, continuous
lasers. Few experiments \cite{Georgiades&Kimble,Georgiades&Polzik}
have performed TPA with coherent, narrowband down-converted light,
demonstrating nonclassical features which appear at very low
powers \cite{Lin_S_89,Lin_E_90,Lin_E_97}, and result from the time
and energy correlations (entanglement) between the down-converted
photon-pairs \cite{Mandel&Wolf,Burnham&Weinberg,Hong&Ou}. At high
power levels (as those discussed in this work) these correlations
vanish, yet similarly nonclassical phase and amplitude
correlations \cite{Mollow&Glauber,McNeil&Gardiner} appear between
the signal and idler beams. Unlike the time and energy
correlations at low powers, these phase and amplitude correlations
cannot be described in the usual form of second-order coherence.
At sufficiently high powers, which greatly exceed the
single-photons regime, broadband down-converted light that is
pumped by a narrowband laser is inherently incoherent, and
exhibits the properties of a broadband thermal noise
\cite{Mandel&Wolf,Scully&Zubairy}; consequently, it is not
expected to be effective at inducing TPA. Nevertheless, it was
shown that at the appropriate conditions, the quantum correlations
within the down-converted spectrum can give
rise to efficient sum-frequency generation \cite{Abram&Raj,QSFG}.\\

In this work we show that down-converted light beams with a
spectral bandwidth that exceeds a certain limit can induce TPA
just like transform-limited pulses with the same bandwidth.
Consequently, the interaction exhibits a sharp, pulse-like
temporal behavior, and can be coherently controlled by
pulse-shaping techniques, even though the down-converted light is
neither coherent nor pulsed. This effect occurs as long as the
transition energy lies within the spectrum of the pump laser that
generated the light; thus, the spectral selectivity of the
interaction is dictated by the narrowband pump laser and not by
the orders of magnitude wider bandwidth of down-converted light
itself. We demonstrate these principles experimentally by inducing
and coherently controlling TPA in atomic Rubidium with
down-converted light, and obtaining results that are practically
identical to those obtained with coherent ultrashort pulses.\\

\begin{figure*} [tbp] \label{fig1}
\begin{center}
\includegraphics[width=16cm]{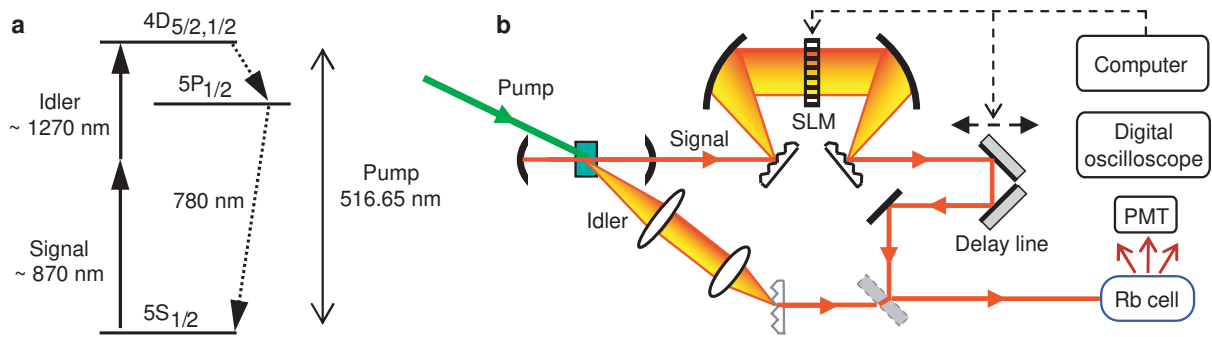}
\end{center}
\caption{Experimental setup for TPA with broadband down-converted
light. (a) Atomic energy levels of Rb, showing the two-photon
transition induced by the broadband signal and idler beams between
the $5S$ and the $4D$ levels, and the resulting decay through the
$5P$ levels. (b) The experimental layout. 3 $ns$ pulses at 516.65
$nm$ were used to pump a 14 $mm$ long BBO crystal that was located
inside a low-finesse resonator. The relative angles between the
crystal axis, the pump laser and the resonator were chosen to
obtain broadband phase matching for non-collinear type-I
down-conversion, with the signal beam propagating in the direction
of the resonator, and the idler beam diverging from it at an angle
of 11$^\circ$. Unlike the signal beam, whose direction of
propagation was set by the resonator, the idler beam had a
wavelength-dependent angular spread, due to the phase matching
conditions in the crystal. Therefore, it was directed through an
imaging system that included a transmission grating which
compensated for this spread. The signal beam was directed through
a pulse shaper and a computer-controlled delay line with 0.1 $\mu$
resolution, and then combined with the idler beam by a dichroic
mirror. The pulse shaper separates the spectral components of the
beam and utilizes a computer-controlled spatial light modulator
(SLM) to introduce the desired spectral phase filter to the light,
before recombining the spectral components back to a single beam.
The combined signal and idler beams then entered the Rb cell. A
photo-multiplier tube (PMT) and a digital oscilloscope were used
to perform a triggered measurement of the TPA-induced
fluorescence.}
\end{figure*}

The underlying principle that enables coherent TPA with broadband
down-converted light is based on the fact that the quantum
interference that governs TPA involves pairs of electromagnetic
modes. Since the excitation of an atomic level with frequency
$\Omega$  may be induced by any two photons with frequencies
$\omega$ and $\Omega-\omega$ , regardless of the exact value of
$\omega$, the final population $p_f$  is proportional to
\cite{PRA60}:

\begin{eqnarray} \label{Ew}
p_f \propto \bigg | \int E(\omega)E(\Omega-\omega)\: d\omega \
\bigg| ^{\:2} ,
\end{eqnarray}

where $E(\omega)$ is the spectral amplitude of the light. As is
obvious from Eq. (\ref{Ew}), it does not matter whether
$E(\omega)$ has a defined phase for every $\omega$, but rather
whether the product $ E(\omega)E(\Omega-\omega)$ has a defined
phase for every $\omega$. Despite the incoherence of each of the
down-converted signal and idler beams, they exhibit exactly this
mutually-coherent phase behavior at frequency-pairs, due to the
inherent phase and amplitude quantum correlations within the
down-converted spectrum \cite{Mollow&Glauber}:

\begin{eqnarray} \label{correlations}
{ \textstyle \frac{1}{\sqrt{ \omega}} }\: E_s(\omega) & \approx &
\textstyle{\frac{1}{\sqrt{\omega_p-\omega}}} \:
E_i^*(\omega_p-\omega) \:,
\end{eqnarray}

where $\omega_p$ is the pump frequency, and $E_s(\omega), \:
E_i(\omega)$ denote the spectral amplitudes of the signal and the
idler beams, respectively. In the case of TPA induced by the
combined signal and idler beams from the same source, these
correlations have a drastic effect when the pump frequency is
equal to the total transition frequency. Combining Equations
(\ref{Ew}) and (\ref{correlations}), letting $\omega_p=\Omega$,
reveals that the random phase of $E_s(\omega)$ is always
compensated by the opposite phase of $E_i(\omega_p-\omega)$. Thus,
the integrand in Eq. (\ref{Ew}) has a constant phase, leading to a
full constructive interference of all the spectral combinations,
exactly as if the interaction was induced by a pair of
transform-limited pulses with the same spectra as the signal and
idler beams. Moreover, this TPA process will be sensitive to
minute delays between the signal and idler beams, to dispersion
and even to pulse shaping, exactly as if it was induced by a pair
of ultrashort pulses. Since this constructive interference occurs
only when the final state energy falls within the spectrum of the
pump laser, the spectral resolution of the TPA process will be
equal to the spectral bandwidth of the pump laser, regardless of
the actual bandwidth of the down-converted light itself. Note that
in the case of a continuous pump laser, this implies a possible
spectral resolution of a few $MHz$ or even less, which is
phenomenally high for an interaction that is induced by light with
a spectral bandwidth
that may be 5-7 orders of magnitude broader. \\

A complete quantum-mechanical analytic calculation
\cite{Calculations}, which takes into account the spectral
bandwidths of the pump laser and of the atomic level, shows that
the TPA signal is composed of two parts, which may be referred to
as the 'coherent TPA' and the 'incoherent TPA'. The coherent TPA
results from the summation of conjugated spectral components, and
indeed can be coherently controlled. The incoherent TPA, however,
results from the summation of all other random spectral
combinations and is a direct result of the incoherence of the
down-converted light; therefore it is unaffected by spectral-phase
manipulations and may be regarded as a background noise, which
limits the equivalence of the down-converted light to a coherent
pulse. The ratio between the coherent term $I^c$ and the
incoherent term $I^{ic}$ can be approximated by :
\begin{eqnarray} \label{ratio}
\frac{I^{c}}{I^{ic}} \approx
\frac{B}{(\gamma_p+\gamma_f)}\frac{n^2+\frac{1}{2\pi}n}{n^2} \: ,
\end{eqnarray}
where $n$ is the spectral average of the mean photon flux, and
$B,\gamma_p,\gamma_f$ are the bandwidths of the down-converted
light, the pump laser and the final state, respectively. This
expression reveals the importance of using spectrally broad
down-converted light. The coherent term becomes dominant only when
the down-converted bandwidth exceeds both the pump bandwidth and
the final level width:
$B>(\gamma_p+\gamma_f)(\frac{n^2}{n^2+\frac{1}{2\pi}n})$. Equation
(\ref{ratio}) shows that the coherent term exhibits a linear
intensity dependence at low powers, as was previously shown
\cite{Lin_S_89,Lin_E_90} and experimentally observed
\cite{Georgiades&Kimble}.\\

Our experimental layout is described in Fig. 1. The pump laser
(Spectra-Physics MOPO-SL laser) was tuned to emit 3 $ns$ pulses
with a bandwidth of 0.04 $nm$ around 516.65 $nm$, which
corresponds to the $5S-4D$ transition in Rb (Fig. 1a). These
pulses pumped a BBO crystal that was located inside a low-finesse
resonator, with the phase-matching conditions chosen to obtain
broadband ($\sim$ 100 $nm$ each), non-collinear signal and idler
beams. The signal beam was directed through a computer-controlled
pulse-shaper, which is normally used to temporally shape
femtosecond pulses and performs as a spectral-phase filter
\cite{Weiner_Shaping}. From the pulse-shaper the signal beam
continued to a computer-controlled delay line, and then was
combined with the idler beam by a dichroic mirror. The combined
beams entered the Rb cell, and the induced TPA was measured
through the resulting fluorescence at 780 $nm$. Our analysis
predicts that the down-converted beams, each 3 $ns$ long with a
peak power of about 1 $MW$ and a spectral bandwidth of about 100
$nm$, should induce TPA with the same efficiency and temporal
resolution as 23 $fs$ pulses with a peak power of about 150 $GW$,
while exhibiting a spectral resolution of 0.04 $nm$.\\
\begin{figure} [t] \label{fig2}
\begin{center}
\includegraphics[width=8.6cm]{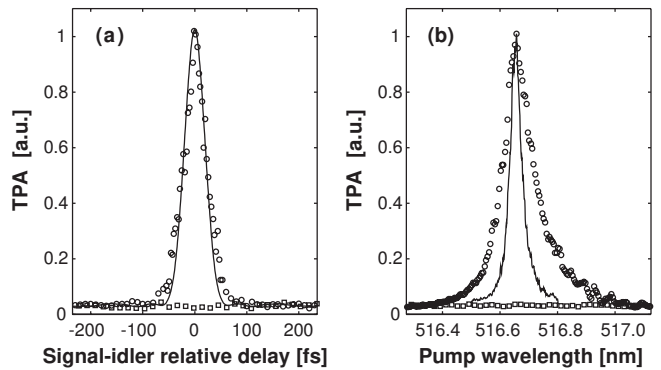}
\end{center}
\caption{Experimental TPA with broadband down-converted light, as
a function of the signal-idler delay and the pump wavelength. The
graphs clearly show a coherent TPA signal that is narrow both
temporally and spectrally, superimposed on an incoherent
background signal, which is insensitive to both pump wavelength
and signal-idler relative delay. In this experiment the pulse
shaper was used only to compensate for the dispersion in our
system. (a) Calculated (line) and experimental TPA with
off-resonance (squares) and on-resonance (circles) pump, as a
function of the signal-idler delay. (b) Experimental TPA at zero
signal-idler delay (circles), and at 100 $fs$ signal-idler delay
(squares), as a function of the pump center wavelength, together
with a typical spectrum of the pump (line). The coherent TPA
signal appears only when the pump is on-resonance with the $5S-4D$
transition, and exhibits a sharp dependence on the signal-idler
delay, exactly as if the interaction was induced by a pair of
coherent, 23 $fs$ pulses with the same spectra as the signal and
idler beams.}
\end{figure}

\begin{figure*} [t] \label{fig3}
\begin{center}
\includegraphics[width=17cm]{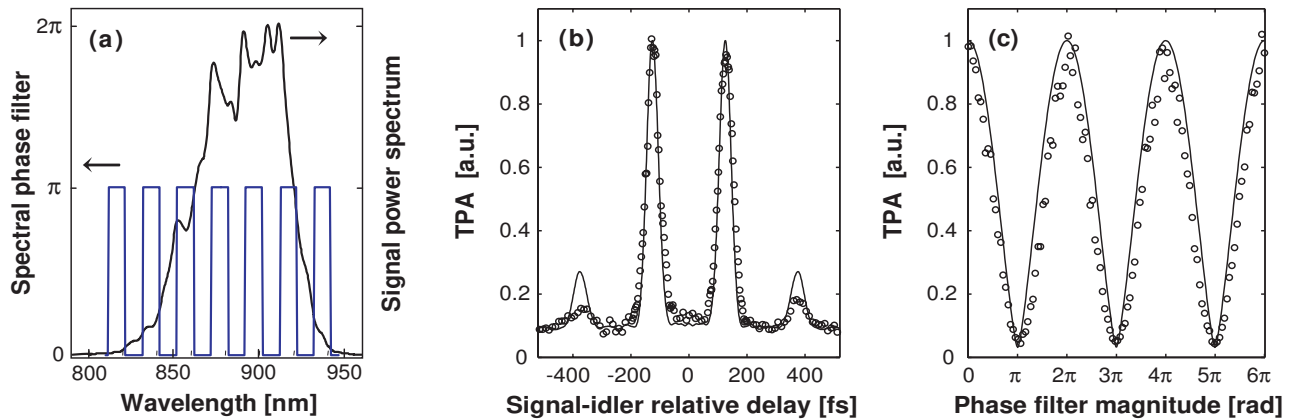}
\end{center}
\caption{Experimental coherent quantum control of TPA with
down-converted light. (a) The square-wave spectral phase filter
applied on the signal beam by the pulse-shaper, together with a
typical power spectrum of the signal. (b) Experimental (circles)
and calculated (line) TPA signal as a function of the signal-idler
delay, with a spectral square-wave phase filter with magnitude
$\pi$ applied to the signal beam. The splitting of the central
peak confirms that the coherent TPA behaves exactly as if it was
induced by a pair of transform-limited pulses, with one of them
temporally shaped by the applied phase filter. (c) Experimental
(circles) and calculated (line) TPA signal as a function of the
magnitude of the square wave spectral phase filter. The graph
shows the periodic annihilation of the coherent TPA at magnitudes
of $\pi, 3\pi , 5\pi$, where a complete destructive quantum
interference occurs ('dark pulse'), and the complete
reconstruction of the TPA signal at amplitudes of $2\pi, 4\pi,
6\pi$  etc, demonstrating complete quantum control over the
coherent TPA process.}
\end{figure*}

To verify these predictions experimentally, we scanned both the
pump wavelength and the relative delay between the signal and
idler beams. Fig. 2 shows the TPA signal versus the signal-idler
relative delay and the pump-wavelength. As is evident, for
off-resonance pump or for large signal-idler delays, only low
efficiency TPA signal is observed, which is insensitive to either
pump wavelength or signal-idler delay. This signal corresponds to
the classical, incoherent TPA background signal. However, when the
pump wavelength was set to the $5S-4D$ transition resonance, an
additional, high efficiency, TPA signal appeared. The sharp
response of this coherent TPA signal to a few femtoseconds delay
between the beams is practically identical to the case of TPA
induced by two coherent, 23 $fs$ pulses. This temporal resolution
is 5 orders of magnitude better than the actual 3 $ns$ temporal
duration of the down-converted light.

Since the peak power of the equivalent transform-limited pulses is
extremely high (150 $GW$), the signal and idler beams had to be
expanded and attenuated in order to avoid complete saturation of
the transition. Unfortunately, we could not attenuate the beams
enough to avoid saturation completely, due to high levels of noise
in our system. As a result, our measurements had to be performed
in a partially saturated regime, where the measured intensity
dependence of the TPA process was less than quadratic, and the
$4D$ level was power-broadened. Thus, the observed 0.12 $nm$
spectral width of the coherent TPA is dictated by the bandwidth of
the pump laser (0.04 $nm$) and the width of the (power-broadened)
$4D$ level ($\sim$ 0.08 $nm$). This spectral resolution is ~2000
times narrower than the total bandwidth of the down-converted light.\\

Finally, we demonstrated coherent quantum control over the
coherent TPA process, in a similar way to coherent control of TPA
with ultrashort pulses \cite{Nature396}. For that we used the
pulse shaper to apply a square-wave phase filter on the signal
spectrum (Fig. 3a). With ultrashort pulses this has the effect of
splitting the pulse temporally to a train of several smaller
pulses. Fig. 3b shows precisely this behavior of the coherent TPA
signal as a function of the signal-idler delay. Fig. 3c depicts
the experimental and theoretical TPA signal at zero delay as a
function of the magnitude of the square wave phase filter. The
results, which are identical to those obtained in coherent control
experiments with femtosecond pulses, show the cyclic transition
between complete constructive and destructive quantum interference
of the different spectral sections, and demonstrate the ability to
fully control the coherent TPA process.\\

Broadband down-conversion of a narrowband pump can be considered
as an optical spread-spectrum source \cite{Viterbi}, which
generates both a broadband white noise key and its conjugate key
simultaneously. A spread-spectrum communication channel can
therefore be established by modulating the phase of one of the
keys at the transmitter and using TPA or sum-frequency generation
to reveal the resulting modulations of the coherent signal at the
receiver. Note that any phase modulations performed on the
incoherent signal beam (or even the mere presence of the signal
beam, in the existence of background noise) cannot be detected
without the idler beam. Furthermore, many such communication
channels can share the same signal and idler beams by assigning a
unique signal-idler delay for each channel, thus creating an
optical code division multiple access (OCDMA) network
\cite{Communication}. Other applications may include nonlinear
microscopy and tomography, where the high efficiency and spatial
resolution of ultrashort pulses can be obtained with continuous,
non-damaging intensities.\\

We wish to thank professor Daniel Zajfman for the loan of the
laser system, Adi Diner for advise and assistance, and professor
Nir Davidson for many helpful discussions.

\end{document}